\documentclass[12pt]{article}
\usepackage{latexsym}
\newcommand{\beq}{\begin{equation}}
\newcommand{\eeq}{\end{equation}}
\newcommand{\beqs}{\begin{eqnarray}}
\newcommand{\eeqs}{\end{eqnarray}}

\newcommand{\tr}{\mathrm{tr}}

\begin{document}
\begin{titlepage}
\vskip 2.5cm
\begin{center}
{\LARGE Open Spin Chains in  Super Yang-Mills at Higher Loops:}\\
{\Large Some Potential Problems with Integrability}\\
\vskip 1.2cm
{\Large Abhishek Agarwal} \
\vskip 1.3cm {\Large Physics Department}\\
{\large City College of the CUNY}\\
{\large New York, NY 10031}\\
abhishek@sci.ccny.cuny.edu \vskip 0.7cm
\end{center}
\vspace{3.14cm}
\begin{abstract}
The super Yang-Mills duals of open strings attached to maximal giant
gravitons are studied in perturbation theory. It is shown that
non-BPS baryonic excitations of the gauge theory can be studied
within the paradigm of open quantum spin chains even beyond the
leading order in perturbation theory. The open spin chain describing
the two loop mixing of non-BPS giant gravitons charged under an
$su(2)$ of the $so(6)$ $R$ symmetry group is explicitly constructed.
It is also shown that although the corresponding open spin chain is
integrable at the one loop order, there is a potential breakdown of
integrability at two and higher loops. The study of integrability is
performed using coordinate Bethe ansatz techniques.
\end{abstract}

\end{titlepage}
\section{Introduction and Summary}
In the present paper we study the mixing of non-BPS giant gravitons
in $\mathcal{N}$=4 SYM beyond the leading order in the 't Hooft
coupling. We pay special attention to open-string like operators
coupled to maximal giant gravitons built out of two of the three
complex scalars $Z,W$ at two loop order. We show by an explicit
computation that the two loop  mixing of such operators can be
described within the paradigm of open spin chains and we construct
the two loop open spin chain. We thus extend the work of Berenstein
and Vazquez\cite{ber-1,ber-2} who had shown that the one loop mixing
of non-BPS giant gravitons can be described by integrable open spin
chains. Our analysis suggests that although the spin chain paragigm
holds beyond the leading order in perturbation theory, the two loop
open spin chain is not integrable.

The discovery of integrability in $\mathcal{N}$=4 SYM\cite{min-1}
has lead to various remarkable insights in the present day
understanding of $\mathcal{N}$=4 SYM and many of its less and
non-supersymmetric cousins. The fact that in 't Hooft's large $N$
limit the mixing of single trace operators of the gauge theory can
be studied and the spectrum of anomalous dimensions of the gauge
theory computed by diagonalizing an associated one  dimensional
quantum spin chain, which, to the extent that it has been possible
to check by explicit computations, appears to be integrable opens up
the possibility of detailed higher loop studies of the gauge theory.
Integrability in superconformal Yang-Mills theory was first
discovered at the one loop level in the seminal paper of Minahan and
Zarembo\cite{min-1} who showed that the mixing of single trace
operators constructed out of the six (real) scalars of the gauge
theory is described by an integrable $SO(6)$ spin chain. This result
has subsequently been generalized and supersymmetrized to
incorporate the mixing of single trace operators involving any of
the SYM fields at the one loop level\cite{beisert-super}. At this
level, the mixing is described by a supersymmetric closed spin
chain, which nonetheless is integrable.

The status of integrability, beyond the one loop order is also very
promising. Higher loop integrability in the gauge theory, which was
first glimpsed in\cite{finiteN} has since been understood in much
grater detail.  It was shown by Staudacher\cite{paba} that higher
loop integrability; at least in some closed sub-sectors of operator
mixing such as the $su(2)$ sector built out of two complex scalars,
the $sl(2)$ sector involving a complex scalar and a single covariant
derivative and the $su(1|1)$ sector involving a complex scalar and a
fermion, can be made manifest as one can write down a set of Bethe
equations whose solutions determine the spectra of anomalous
dimensions of the gauge theory and also produce the factorized
scattering matrix of the closed spin chains that describe these
sectors. The insights gained from the Bethe equations that determine
the spectra in these various closed 'two-component' sectors have
been utilized by Beisert and Staudacher in\cite{bs-all-loops} where
a remarkable all loop Bethe ansatz was proposed. Assuming the
correctness of this Bethe ansatz, one is lead to the extremely
pleasant situation where the all-loop large $N$ spectrum of
anomalous dimensions of all single trace operators of $N=4$ SYM is
described by an underlying integrable quantum spin chain. Though,
one does not yet know the Hamiltonian of this complete
supersymmetric spin chain one can write down the all loop
Hamiltonian in some less supersymmetric sectors. For example in the
$su(2)$ sector, an all loop Hamiltonian that reproduces the proposed
$su(2)$ Bethe ansatz and the appropriate dispersion relations can be
written down in terms of a strong coupling expansion of the Hubbard
model\cite{hubbard}. Thus, as far as the large $N$ dynamics of
single trace operators of the gauge theory is concerned, one has
many indications that it is governed by an integrable (closed) spin
chain.

Apart from single trace operators, the gauge theory also contains
baryonic operators, whose engineering dimensions are of order $N$.
These operators are constructed out of determinants and
sub-determinants rather than traces\cite{bal-1,jevicki-1,koch-1}.
Such baryonic operators are of particular interest in the context of
the AdS/CFT correspondence as they provide one with the gauge theory
dual of non-perturbative D brane excitations i.e the so called giant
gravitons of the dual string
theory\cite{witten-giant,susskind-giant,myers-giant,aki-giant,grisaru-giant}.
The simplest example of a giant graviton is provided by a
determinant operator constructed out of one of the three complex
scalar fields $Z$ \beq\epsilon ^{i_1\cdots i_{N}}\epsilon
_{j_1\cdots j_{N}}Z^{j_1}_{i_1}\cdots Z^{j_N}_{i_N}.\eeq This
operator being built out of only one of the complex scalars is a BPS
operator and does not undergo renormalization. The gravitational
interpretation of this operator is that of a maximal D3-brane that
saturates the angular momentum
bound\cite{susskind-giant,myers-giant,aki-giant,grisaru-giant,das-giant}.
Non-BPS excitations in the background of  a giant graviton would
correspond to open string excitations in the presence of the brane.
The gauge theory provides us with a simple enough prescription for
constructing the boundary  theory dual of the open string degrees of
freedom. The open string ending on the D3 brane can be built by
replacing one of the $Z$s in the determinant with a matrix product
of local super-Yang-Mills fields\cite{bal-2,ber-1,ber-2,ber-3}. A
typical non-BPS $su(2)$ operator would look like  \beq\epsilon
^{i_1\cdots i_{N}}\epsilon _{j_1\cdots j_{N}}Z^{j_1}_{i_1}\cdots
Z^{j_{N-1}}_{i_{N-1}}\left(WWZZZWZWZ\cdots
ZZWZW\right)^{j_{N}}_{i_{N}}.\label{typical}\eeq Such operators, by
virtue of not being protected by any non-renormalization theorems,
would undergo operator mixing and hence have non-zero anomalous
dimensions. The spectrum of anomalous dimensions of such operators
gives one the gauge theory dual of the open string spectrum.
Computing the spectrum of anomalous dimensions of such non-BPS
maximal giants is the problem that we shall concern ourselves with
in the present paper. As with various other issues concerning the
AdS/CFT correspondence, it is important to analyze the problem in
the large $N$ limit before addressing the problem in all its
generality.

As it stands, it is not at all obvious that the computation of
anomalous dimensions of giant graviton like operators simplifies in
any way in the large $N$ limit. Processes that naively appear to be
sub-leading order in $\frac{1}{N}$ may in fact turn out to be of
order one as various $\frac{1}{N}$ suppressions may be compensated
for by the $O(N)$ size of the operators. The fact that such
counter-intuitive turn of events can come about in the study of
Yang-Mills theories is rather well known from the study of the
$\mathcal{N}$=4 SYM in the plane wave limit. There too one  studies
long operators that have large bare conformal dimensions and
R-charge, J, and one has the result that if the ratio
$\frac{J}{N^2}$ is held fixed then the usual notions of planarity do
not apply any more\cite{beisert-bmn}. Non-planar contributions turn
out to be of $O(1)$ in that limit. The study of operator mixing can
still be reduced to a quantum mechanical problem, though the
underlying quantum mechanics that one is faced with is not that of a
quantum spin chain but that of a non-local many-body problem. Thus
one of the  pleasant features of the gauge theory, namely
integrability is no longer present in this novel-large $N$
limit\footnote{ In various other scenarios as well, where one takes
into account non-planar splitting and joining of strings the notion
of integrability, if any, appears to be far from obvious. We shall
refer to \cite{split-1, split-2} for some recent developments in
this direction.}.

Though we shall not aspire to construct any new large $N$ limit
where some parameter (other than $N$, such as the R-charge) is
allowed to approach a critical value, and work completely within the
framework of 't Hooft's large $N$ limit, the baryonic operators that
we shall work with are indeed of order $N$. Thus we shall keep the
lessons learnt from the study of the plane wave limit as a valuable
caveat. From the studies of the dynamics of non-BPS giant gravitons
of the gauge theory performed so far, the status of integrability is
ambiguous. It was first shown in\cite{bal-2} that, at least at the
one loop level, the anomalous dimensions of non-BPS giants are not
of order one i.e although these operators have large bare conformal
dimensions, their anomalous dimensions are small. Subsequently, more
detailed  one-loop analysis of non-BPS operators involving all six
(real) SYM scalars was carried out in\cite{ber-1}. In that paper,
one loop mixing of operators such as\beq\epsilon ^{i_1\cdots
i_{N}}\epsilon _{j_1\cdots j_{N}}Z^{j_1}_{i_1}\cdots
Z^{j_{N-1}}_{i_{N-1}}\left(\phi ^{i_1}\cdots \phi
^{i_l}\right)^{j_{N}}_{i_{N}},\eeq where $\phi ^i$s can be any of
the six scalars, was carried out. It was shown that the computation
of one-loop anomalous dimensions of such operators can be reduced to
the problem of diagonalizing the Minahan-Zarembo  $so(6)$ quantum
spin chain, but with open boundary conditions. The open boundary
conditions emerge naturally in the problem as the operators of
interest are not traces, and the interaction of the scalar fields
that form the chain(the matrix product of the $\phi$ fields in the
above operator) with the $Z$ fields forming the brane lead to the
open boundary conditions. The outcome of the analysis performed
in\cite{ber-1} was the discovery that the open boundary conditions
do not spoil the integrability of the Minahan-Zarembo spin chain. It
was shown in that paper that the bulk and boundary $S$ matrices for
the relevant $so(6)$ spin chain satisfy the Yang-Baxter and
reflection algebras: thus a necessary condition for integrability is
indeed satisfied at the one-loop level. However, already at the one
loop level, the situation is different if one considers non-BPS
non-maximal giants, which correspond to operators such as
\beq\epsilon ^{i_1\cdots i_ki_{k+1}\cdots i_{N}}_{i_1 \cdots
i_kj_{k+1}\cdots j_{N}}Z^{j_{k+1}}_{i_{k+1}}\cdots
Z^{j_{N-1}}_{i_{N-1}}(w)^{j_{N}}_{i_{N}}, \eeq with $k>1$. $w$
stands for the matrix product formed out of two complex SYM scalars
i.e. it is the $su(2)$ spin chain.  It was shown in\cite{ber-2} that
the mixing of non-maximal giant gravitons of the above type formed
out of two complex scalars can be formulated in the language of
quantum spin chains, at least at the one loop level. The spin chain
is nothing but the celebrated Heisenberg spin half model, but the
boundary conditions generated by non-maximal giants are of a novel
sort. The boundary interactions result in the spin chain being a
dynamical one, i.e, the boundary interactions lead to fluctuations
in the length of the chain. Though, it was possible to find the
exact ground state of the spin chain in\cite{ber-2} whether or not
the chain is integrable is still an open issue\footnote{The
dynamical nature of the spin chain does not by itself rule out
integrability. Even in the sector of closed chains e.g in the closed
$su(2|3)$ sector dynamical chains do appear\cite{dynb} and they
appear to be integrable\cite{dyna}. However, it is not known if the
dynamical spin chains of interest in the sector of giant gravitons
of the gauge theory enjoy similar integrable features.}. Thus,
integrability in $N$=4 SYM is not at all obvious once one ventures
outside the realm of single trace operators.

Given this status quo, it is probably a reasonable goal to try and
examine whether the spin chain interpretation of mixing of non-BPS
giant gravitons holds beyond the one loop level and if it does it is
of course interesting to investigate whether or not they can be
diagonalized by Bethe ansatz techniques that have been so successful
in the study of higher loop integrability of single trace/closed
spin chain operators. In the present paper we shall take a step in
that direction. The organization of the paper and the principal
results reported in various sections are as follows.

In the next section we shall establish that the dynamics of non-BPS
excitations in the background of a maximal giant graviton can be
studied within the paradigm of open spin chains at the two loop
level. In other words the mixing of non-BPS maximal giants will be
shown to be described by open spin chains at the second order in the
't Hooft coupling. The analysis will be performed in two steps. We
shall  first resolve various combinatorial issues to work out the
proper normalizations for various determinant and sub-determinant
like operators. The guiding principle used in this exercise is that
the various operators are required to have a free field two point
function of order one. Once the combinatorial issues are settled, we
shall proceed to study the mixing of these non-BPS giant gravitons.
To that end we shall use the results of Beisert et al\cite{finiteN}
where the two loop dilatation operator for the $su(2)$ sector of the
gauge theory was computed at finite $N$. Isolating the $O(1)$ part
of the result of the action of the finite $N$ dilatation operator on
the properly normalized operators gives us the details of mixing of
non-BPS giants at large $N$. We shall be able to show that at the
two loop level  non-BPS giants of the type(\ref{typical}) are closed
under action of the dilatation operator. We shall also be able to
interpret the process of mixing of such operators within the
language of open spin chains; the open spin chain being the matrix
product of various $su(2)$ SYM fields that are contracted with the
last pair of indices of the epsilon tensors in(\ref{typical}). We
shall also be able to show that no new boundary conditions are
generated at the two loop level. Thus the boundary conditions are
the same as the ones discovered by Berenstein and collaborators
in\cite{ber-1}, which amount to the fact that the first and the last
fields in the open chain(\ref{typical}) cannot be $Z$s. These
boundary conditions emerge out of the realization that if a $Z$
field is placed at either end of the open spin chain, then the
operator factorizes into a determinant and a trace, which is lower
order in $\frac{1}{N}$. This fact may most easily be seen by going
to a basis in which $Z$ is diagonal. Thus the boundary conditions
for the problem at hand are not exactly the so called free boundary
conditions, which would have corresponded to the case of the first
and last spins being allowed to take any one of the two allowed
values, but they are very similar.

Having a spin chain interpretation of the  mixing of non-BPS giants
at two loops is a gratifying outcome. The one loop open spin chain
was already shown to be integrable in\cite{ber-1}, and our analysis
suggests that the study two loop operator mixing can be regarded as
the analysis of a perturbation to an integrable open spin chain. The
obvious question that comes to mind is whether or not the two loop
spin chain Hamiltonian can be diagonalized by an appropriate Bethe
ansarz. That, unfortunately does not appear to be the case. Before
indulging in further details, it is probably worth emphasizing that
even in the absence of integrability in the sense of the absence of
a Bethe ansatz and a factorized $S$ matrix, the fact that the two
loop operator mixing can be described by open spin chains is a major
simplification in the study of the open string/giant graviton sector
of the gauge theory. That is so because, as far as the computation
of the spectrum of anomalous dimensions of operators that correspond
to spin chains of finite length is concerned, the spin chain
interpretation always reduces the problem to the process of
diagonalization of a finite dimensional matrix, which can be
achieved numerically. However, the lack of a two loop  Bethe ansatz
is a major drawback if one intends to study long coherent open
string states constructed from the gauge theory and compare with the
world-sheet analysis as was done in\cite{bal-2}.

Having established the open spin chain description of the mixing of
non-BPS giants, we shall devote the rest of the paper to a careful
analysis of the integrability or the lack thereof of the two loop
open spin chain. Although integrability of a given spin chain, even
in a perturbative sense, is in general quite hard to establish, we
shall take the pragmatic approach of Staudacher\cite{paba}  and
declare the spin chain to be integrable if one can successfully
write down a set of Bethe equations for the spin chain. In the
approach pioneered in\cite{paba}, the basic idea was to study the
two magnon scattering problem, and find the appropriate two body $S$
matrix. Assuming factorized scattering, the two solution of the two
body problem was then used to find the exact (in a perturbative
sense) eigenstates, i.e the Bethe eigenstates  of the general many
body problem. An explicit construction of the scattering eigenstates
of the $su(2)$ dilatation operator and the corresponding Bethe
equations, accurate to the third order in the 't Hooft coupling was
also worked out by Staudacher in the same paper\cite{paba}. As far
as the closed chains of the $su(2)$ sector is concerned, the main
complexity lay in the construction of the two magnon scattering
matrix and the Bethe equations describing the two magnon scattering.
As long as there is no global obstruction to the construction of the
two particle $S$ matrix, one could assume factorized scattering and
lift the solution of the two body problem to the case of an
arbitrary number of magnons. One could then carry out further checks
to ascertain that the solutions of the general many body problem do
in fact reproduce the correct spectrum. In the $su(2)$ sector, for
instance, the spectrum obtained from the Bethe ansatz\cite{paba}
does match the spectrum obtained from the numerical lattice
simulations of the corresponding spin chain\cite{virial}. In a
sense, the notion of integrability employed in this approach is the
factorized nature of the scattering matrices for  three or more
magnon scattering processes.

However, in the presence of boundaries, there is a second notion of
three body scattering, which is present even in the analysis of the
two magnon problem. The presence of the boundary can lead to three
body processes even if one is dealing with the dynamics of two
magnons. Integrability in the presence of boundaries hence implies
that these three body interactions take place in a way that is
compatible with the separation of the scattering matrix into a bulk
and a boundary part. Thus, even from the point of view of
Staudacher's PABA apprach, the criterion of integrability is more
stringent in the presence of non-periodic boundary conditions. We
shall be able to demonstrate, in an explicit fashion that such three
body processes, though absent at the one loop level, do start
contributing at $O(\lambda ^2)$. The algorithm that we employ to do
this is the following one. We compute the bulk and boundary $S$
matrices from the study of the asymptotics of the two magnon
problem. In other words, the study of the magnons scattering off
each other far away from the boundaries gives us the bulk two body
$S$  matrix of the spin chain and the situation when one of the
magnons scatters off the boundary (while being far enough away from
the other magnon) gives us the boundary $S$ matrix. From the
knowledge of these scattering matrices, we can write down a set of
Bethe equations. We then proceed to show that these Bethe equations
follow from the requirement for a general ansatz for the two
particle wave function to be an eigenstate of the spin chain
Hamiltonian. Finally we demonstrate explicitly that the two body
wave function that leads to the Bethe equations is not an eigenstate
of the Hamiltonian. We interpret this to be evidence in favor of a
breakdown of higher loop integrability in the open string sector of
the gauge theory. Though the Bethe equations do not appear to be
solutions of the two body Schrodinger equations, we can nevertheless
utilize them to construct some exact eigenfunctions of the
Hamiltonian. This construction is discussed at the very end of the
third section.

From the point of view of the spin chain, we can also understand the
potential breakdown of integrability in another way. The two loop
dilatation operator that describes the mixing of non-BPS giants is
the same as the one for closed chains as long as one stays away from
the boundaries, i.e the bulk dilatation operator can be embedded in
the Inozemtsev spin chain\cite{ino-1}. To the best of our knowledge,
the only boundary conditions, other than the periodic ones, that are
compatible with the integrability of the Inozemtsev chain are
provided by reflective boundaries (see for
example\cite{serban-open,bn-open, xing-open}). The boundary
conditions that emerge out of the gauge theory analysis are not of
that type, and as mentioned earlier, they are similar in spirit to
the free boundary conditions. The Inozemtsev chain is not known to
be integrable in the presence of free boundary conditions, and hence
it is perhaps not surprising from this point of view that the two
loop dilatation operator in question in the present paper does not
appear to admit a consistent Bethe ansatz.

\section{Various Operators:}
The generic class of operators of interest to us are the ones that
describe non-BPS excitations in the background of a maximal giant
graviton. As mentioned earlier in the introduction, maximal BPS
giants are protected operators formed out of determinants of a
single complex scalar $Z$, and they define the vacuum state for the
sector of interest to us.\beq|0> = \epsilon_{i_1\cdots
i_N}^{j_1\cdots j_N}Z^{i_1}_{j_1}\cdots Z^{i_N}_{j_N}\eeq It is
understood that\beq\epsilon_{i_1\cdots i_N}^{j_1\cdots j_N} =
\epsilon_{i_1\cdots i_N}\epsilon^{j_1\cdots j_N}.\eeq A typical open
string excitation corresponds to a state such as
 \beq|k;Z^m;w_{1},\cdots
w_{N-k-m}>=\epsilon ^{i_1\cdots i_ki_{k+1}\cdots i_{N}}_{i_1 \cdots
i_kj_{k+1}\cdots j_{N}}Z^{j_{k+1}}_{i_{k+1}}\cdots
Z^{j_{k+m}}_{i_{k+m}}(w_1)^{j_{k+m+1}}_{i_{k+m+1}}\cdots(w_{N-k-m})^{j_{N}}_{i_{N}}
\eeq The $w_i$s are matrix products of local Yang-Mills fields i.e
they are the open string degrees of freedom. In the above operator,
the indices of the epsilon tensor ($k$ of them) which are contracted
with each other indicate the departure from the maximal $R$ caharge
bound, while, replacing a certain number of the $Z$ fields with the
open strings, $w_i$, signal the departure form the BPS bound. As has
been argued and demonstrated in\cite{ber-3}, operators of the above
kind do satisfy orthonormality conditions and it is possible to
build a large $N$ Hilbert space using these open string degrees of
freedom once the states are properly normalized. Rather than review
the detailed combinatorial work presented in\cite{ber-3}, which
deals with the generic class of operators to which the above example
belongs we shall focus on the particular operators of interest to
us. We shall be interested in operators of the type \beq
|Z^{N-1},w>\eeq which gives us the gauge theory description of a
single open string ending on a maximal giant graviton. \beq w^i_j =
\left(w_1 w_2 \cdots w_L\right)^i_j.\eeq Each of the 'bits' of the
open string $w_i$ can take on two values $Z$ and $W$.
\subsection{Normalization:} To be able to develop a consistent large
$N$ expansion one needs to normalize the operators in a way that
lends itself to a well defined large $N$ limit. We shall chose to
normalize the operators by the free-field two point functions. It
may be seen by a straightforward computation that
\beq\left<\bar{Z}^{N-1}(x),\bar{w}(x)|Z^{N-1}(0),w(0)\right>\sim
[(N-1)!]^3N^{L+1}\frac{1}{|x|^{N+L-1}}\eeq Hence, it is reasonable
to normalize the  states as \beq
\frac{1}{\sqrt{[(N-1)!]^3N^{L+1}}}|Z^{N-1},w>\eeq As mentioned in
the introduction, with this normalization one is  led immediately to
dirichlet boundary conditions for the open spin chain $w$. It may be
easily verified that if either $w_1$ or $w_L$ is equal to $Z$, then
the state factorizes into a product of a determinant and a single
trace\cite{ber-1,ber-3}. With our chosen normalization, such a state
would turn out to be sub-leading order in the large $N$ limit. Thus,
one has the Dirichlet boundary conditions that, neither the first or
the last impurity fields can be of the $Z$ type.

\subsection{The Open Spin Chain:}The two loop dilatation operator of
$\mathcal{N}$=4 SYM in the $su(2)$ sector is\cite{finiteN} \beq D =
D_0 + \frac{\lambda }{N} D_1 + \frac{\lambda ^2 }{N^2} D_2
+O(\lambda ^3)\eeq where \beq D_0 = \tr\left(A^{\dagger Z}A_Z
+A^{\dagger W}A_W\right)\nonumber\eeq \beq D_1 = -\tr [A^{\dagger
Z},A^{\dagger W}][A_Z, A_W] \nonumber \eeq \beqs D_2
&=&\tr(\frac{1}{2} [A_Z, A_W][A^{\dagger
Z},[A_Z,[A^{\dagger Z},A^{\dagger W}]]]\nonumber \\
& & +\frac{1}{2}[A_Z, A_W][A^{\dagger W},[A_W,[A^\dagger _Z,
A^\dagger _W]]]\nonumber
\\
& & + [A^{\dagger Z},A^{\dagger W}][A_Z, A_W]),\eeqs  and \beq
\lambda = \frac{g^2_{YM} N}{8\pi^2}.\eeq It is understood, that one
has mapped the local composite ($su(2)$) operators to states of the
matrix model Hamiltonian. For example,\beq Tr(ZZWW\cdots
W)\Longleftrightarrow Tr(A^{\dagger Z}A^{\dagger Z}A^{\dagger
W}\cdots A^{\dagger W})|0>,\eeq and that operator mixing is mapped
to the action of the dilatation operator, thought of as a matrix
model Hamiltonian, on various states like the one depicted above.
When the dilatation operator acts a giant graviton like state such
as $|Z^{N-1},w>$ the following possibilities may occur.\\
{\bf (1:)} All the annihilation operators of $D_1$ and $D_2$ may act on $w$.\\
{\bf (2:)} Some of the annihilation operators may act on $w$ and
others on the $Z$ fields that form the brane.\\
In the first instance, one has the standard result that the action
of the annihilation operators on consecutive fields present in $w$
can be interpreted  as the action of a quantum spin chain
Hamiltonian on a spin chain state\cite{finiteN,abhi-1}. The action
of the annihilation operators on non-consecutive fields can be shown
to be sub-leading order in $\frac{1}{N}$. Such actions  always spit
up the spin chain into multiple chains,  thus transforming the state
with a single string excitation in the D3 background to one
involving multiple open string excitations. Such states are mutually
orthogonal (see\cite{ber-3} for the combinatorial details). Hence
they can be ignored for the purposes of large $N$ computations. The
spin chain that results from the 'bulk' action of the dilatation
operators on consecutive fields present in $w$ is the same as the
one that one has in the study of single trace
operators\cite{finiteN}, only the boundary conditions are different.
As mentioned previously, the boundary conditions consist of the
condition that neither the first or the last fields of the spin
chain can be a $Z$.

When the dilatation operator acts on two $Z$ fields from the brane,
the result can be easily shown to be sub-leading order in
$\frac{1}{N}$. However, special care is required to analyze the
result of the action of the dilatation operator on two consecutive
fields in $w$ and a single $Z$ field from the brane. A careful
analysis shows that there are no $O(1)$ contributions from such
actions of the dilatation operator. Indeed, all the terms so
generated are given below along with the leading powers of
$\frac{1}{N}$ by which they are suppressed. In the following list,
$w_1$ stands for a single field, $w_2$ represents a generic 'word'
formed out of two fields and $w_c$ is a word of length $|w|-2$:
$|w|$ being the length of $w$. The second set of states multiplying
the fifth and lower terms in the table are single trace states which
result upon the action of the dilatation operator on the giant
graviton state.

\begin{center}

\begin{tabular}{|c|c|}
  $|Z^{N-2},w_1,w_cw_2>$ & $\frac{1}{\sqrt{N}}$\\
  $|Z^{N-2},w_3,w_c>$ & $\frac{1}{\sqrt{N}}$\\
  $|Z^{N-2},w_2,w_1w_c>$ & $\frac{1}{\sqrt{N}}$\\
  $|1,Z^{N-2},w_3w_c>$ & $\frac{1}{\sqrt{N}}$\\
  $|1,Z^{N-2},w_c>|w_3>$ & $\frac{1}{\sqrt{N}}$\\
  $|1,Z^{N-2},w_2w_c>|w_1>$ & $\frac{1}{\sqrt{N}}$\\
  $|1,Z^{N-2},w_1w_c>|w_2>$ & $\frac{1}{\sqrt{N}}$\\
  $|Z^{N-2},w_2,w_c>|w_1>$ & $\frac{1}{N}$\\
  $|1,Z^{N-2},w_2w_c>|w_1>$ & $\frac{1}{\sqrt{N}}$\\
  $|Z^{N-2},w_1,w_c>|w_2>$ & $\frac{1}{\sqrt{N}}$\\
\end{tabular}

\end{center}
We can now write down the formula for the two loop large $N$ open
spin chain. As mentioned in the preceding discussion the two loop
Hamiltonian for  the open string sector is the same as the one for
single trace operators, the only thing that is is different is the
boundary condition. To incorporate the boundary conditions in the
formula for the Hamiltonian we shall use the operators $Q_i(Z)$,
which were introduced in\cite{ber-1}. These operators are defined by
their actions on the spin chain states as
foillows: \beqs Q_i^Z|\cdots Z\cdots> &=&0\nonumber\\
Q_i^Z|\cdots W\cdots> &=& |\cdots W\cdots>. \eeqs In the above
equations, $Z,W$ are located at the 'i'th lattice site. i.e. the
operator $Q_i(Z)$ checks if the 'i'th spin is a $Z$. If it is so
then it annihilates the entire state. Using these operators, the
open spin chian Hamiltonian can be written as \beqs H &=&
\sum_{l=1}^{J-1}(1 -4g
)Q^Z_1Q^Z_J(I-P_{l,l+1})Q^Z_1Q^Z_J\nonumber \\
& & +g\sum_{l-1}^{J-2}Q^Z_1Q^Z_J(I-P_{l,l+2})Q^Z_1Q^Z_J\nonumber
\eeqs To get to the above form, we have divided out the formula for
the Hamiltonian by $\lambda $ and used \beq g =\frac{\lambda
}{2}.\eeq It is of course assumed that the length of the chain is
$J$ and the Hamiltonian acts on all the spin sites. Since the first
and the last sites cannot have $Z$ fields, one could write down the
Hamiltonian is a slightly different form as well. We could take a
state of the generic form \beq|W\cdots W>,\eeq where one could have
impurity ($Z$) fields inserted between two $W$ fields. In the
previous formulation, the spin chain included the boundary fields,
but we could just as well modify the definition of the spin chain to
include only the fields contained inside the two boundary $W$
fields, and write the effect of the interaction of the boundary
fields with the spin chain separately. In other words, we shall
define the second field of $w$ to be first site of the spin chain
and the penultimate field of $w$ to be the last site. Thought of in
this way, the spin chain Hamiltonian reads as\beqs H =
\sum_{l=1}^{L-1}(1-4g)(I-P_{l,l+1}) +
g\sum_{l=1}^{L-2}(I-P_{l,l+2})\nonumber \\
(1-4g)q^Z_1 + gq^Z_2 + (1-4g)q^Z_L + gq^Z_{L-1} \label{dil2}\eeqs It
is to be understood that\beqs q^Z_i = (1- Q^Z_i),\eeqs i.e $q^Z_i$
checks if the field at the $i$th site is $Z$, and if it is so, it
acts as the identity operator. In case this condition is not
satisfied, $q^Z_i$ annihilates the spin chain.  $L$ is now to be
regarded as the effective length of the spin chain i.e the number
for fields between the boundary $W$ fields i.e. \beq L = |w|-2.\eeq
\subsection{Bethe Ansatz:}
Let us begin by deriving the one loop Bethe ansatz for the open spin
chain. Short ranged spin chains of the Heisenberg type with various
boundary conditions have been solved using both the coordinate and
algebraic Bethe ansatz techniques in the
past\cite{open-1,open-2,open-3}. In the present paper, we shall
adhere to the coordinate Bethe ansatz approach as it generalizes
easily to higher loops. An excellent discussion of coordinate Bethe
ansatz solutions to short ranged spin chains can be found
in\cite{open-1,open-2}, and we shall adhere to the approach
presented in these papers in this present analysis.

The one magnon state $|\Psi_1>$ can be taken to be of the form:\beq
|\Psi_1> = \sum_{x=1}^{L}\Psi (x)|x>. \eeq If one takes the state to
be a superposition of incoming and outgoing plane waves, with\beq
\Psi (x) = A(k) e^{ikx} - A(-k)e^{-ikx}, \eeq then it is quite easy
to see that the state is an eigenstate of the Hamiltonian with
energy \beq E(k) = 4\sin ^2\left(\frac{k}{2}\right).
\label{spec-1}\eeq It is also straightforward to verify that for the
plane wave state to be an eigenstate, the following boundary
conditions, which are implied by scattering at the boundary, must
hold. \beqs \Psi (0) &=& 0
\nonumber \\
 \Psi (L+1) &=& 0. \eeqs These conditions can also  be written down as\beqs
 \alpha (-k)A(k) -  \alpha (k)A(-k) = 0\nonumber \\
 \beta (k)A(k) -  \beta (-k)A(-k) = 0.\label{comp1}\eeqs
 The above conditions lead to the following compatibility condition\beq
 \alpha (k)\beta (k) =  \alpha (-k)\beta (-k)\eeq
 with $\alpha, \beta $ given by
\beqs \alpha (-k) &=&  1\nonumber \\
\beta (k) &=& e^{i(L+1)k}. \eeqs The solution to the compatibility
condition sets the allowed range of values for the momenta to be\beq
k = \frac{n\pi }{L +1}.\label{mom-1}\eeq This is the complete
solution to the one magnon scattering problem, which gives us the
one-magnon spectrum(\ref{spec-1},\ref{mom-1}) and the boundary
scattering matrix $\frac{A(k)}{A(-k)}$.

Having determined the boundary $S$ matrix by solving the one-magnon
problem, one can now proceed to study the two magnon scattering
problem. The two magnon state can be taken to be of the form\beq
|\Psi_2> = \sum_{x<y}\Psi(x,y)|x,y>\eeq with \beq\Psi (x,y) = \sum
_p \sigma (p)A(k_i,k_2)e^{i(k_1x_1 +k_2x_2)}.\eeq The above sum is
over all permutations and negations (p) of the momenta. $\sigma (p)
= \pm 1$ and it changes sign when the signs of when one of the signs
of the moments are changed or when the momenta are permuted in
pairs. This two magnon state is an eigenstate of the spin chain
Hamiltonian provided the following conditions are satisfied. \beqs
A(k_1,k_2) s(k_1,k_2) - A(k_2,k_1)s(k_2,k_1) =0\nonumber
\\A(k_1,k_2)\alpha (-k_1) -
A(-k_1,k_2)\alpha (k_1) =0\nonumber \\
A(k_1,k_2)\beta (k_2) - A(k_1,-k_2)\beta (-k_2)
=0\label{one-loop}\eeqs In the first equation \beq s(k_1,k_2) =
1-2e^{ik_2} + e^{i(k_1+k_2)}.\eeq The three equations above require
some explanation. The first of the three equation comes from the
requirement that two body scattering, that takes place when the
impurities appear next to each other, happens in a way such that the
state remains an eigenstate of the Hamiltonian. The other two
equations follow from scattering at the boundary, i.e by considering
the scenario that one of the particles is at the boundary and by
requiring the coefficient of this state to be such that it still
remain an eigenstate of the Hamiltonian. In a sense we can see,
directly from these equations that the boundary conditions are
compatible with integrability at the one loop level. That follows
from the observation that the boundary scattering equations involve
only one of the two momenta. In other words, the boundary does not
introduce any new 'three body' interactions. For example, if one
considered the special case when the two particles sit next to each
other at the boundary, there could in principle have been potential
three body interactions; the third body being the boundary. However,
the above conditions tell us that this is not the case at the one
loop level, and the two boundary equations are all that one needs
for {\it all } positions of the second impurity. We shall soon see
that this is {\it not } the case at the two loop level.

The situation is more transparent in  position space. For any pair
of well separated lattice sites away from the boundary, \beq H |x,
y(>x+1)> = 4|x,y> - |x-1,y> - |x+1,y> - |x,y-1> - |x,y+1>\eeq With
the choice of the pane wave ansatz, and this action of the
Hamiltonian, it is clear that the coefficient of $|x,y(>x+1)>$, once
the Hamiltonian has acted on it, is
$(4\sin^2\left(\frac{k_1^2}{2}\right) +
4\sin^2\left(\frac{k_2^2}{2}\right))\Psi(x,y)$. However, for the
coefficient of $|x, y(=x+1)>$ to come out right, one has to impose
the condition\beq 2\Psi(x,x+1) - \Psi(x+1,x+1) - \Psi(x,x) =0\eeq
which leads to the scattering condition; the first equation in
(\ref{one-loop}). Once this equation is satisfied, the effects of
the boundary  impose the further constraint\beqs \Psi(0,y) &=&0\nonumber \\
\Psi(x,L+1) &=& 0\eeqs which translate into the second and third
equations in (\ref{one-loop}) respectively. What is notable is that
once the two particle scattering equations are satisfied, for some
choice of $\Psi(x,y)$, the boundary conditions on $\Psi(x,y)$ are
the same as the one magnon case. For instance, the left boundary
condition (the first of the two equations above) is the same as the
condition satisfied by $\Psi(x)$ in the one-magnon problem for all
values of $y$. Most importantly, the special situation when $x=1,
y=2$ does not introduce any new conditions on the wave function.
Thus the bulk and boundary scattering problems separate out. This is
nothing but an explicit (and perhaps pedagogical) way of realizing
that the boundary conditions  are compatible with integrability of
the spin chain.

Having the three equations at out disposal, we can now derive the
Bethe equations that determine the momenta. As in the one magnon
case, the Bethe equations are derived as the compatibility
conditions implied by the three equations. Various ways of
expressing $A(k_1,k_2)$ in terms of $A(k_2,k_1)$ lead to the
requirement\cite{open-2}\beq \frac{\alpha (k_1)\beta (k_1)}{\alpha
(-k_1)\beta (-k_1)} = \frac{S(-k_1,k_2)}{S(k_1,k_2)},\eeq where \beq
S(k_1,k_2) = s(k_1,k_2)s(k_2,-k_1)\eeq There is a similar equation
where $k_1$ is replaced by $k_2 $ and vice-versa. These results can
be generalized to the case of an arbitrary number of magnons. In the
case of $m$ magnons the Bethe equations read as\beq \frac{\alpha
(k_i)\beta (k_i)}{\alpha (-k_i)\beta (-k_i)} = \prod_{j(\neq
i)=1}^m\frac{S(-k_i,k_j)}{S(k_i,k_j)}.\eeq
\section{The Two Loop Analysis:}
Bethe ansatz for the $su(2)$ open spin chain decribed in the
previous section is essentially implied in the analysis carried out
in\cite{ber-1} where the one loop bulk and boundary $S$ matrices of
the full $so(6)$ spin chain were computed. The above analysis
reconfirms and reproduces the $su(2)$ sector of the  results
reported in\cite{ber-1}from a coordinate Bethe ansatz point of view
and also sets up the techniques and conventions that will be applied
to the two loop analysis, to which we shall devote the rest of the
paper.
The two loop large $N$ dilatation operator is\beqs H =
\sum_{l=1}^{L-1}(1-4g)(I-P_{l,l+1}) +
g\sum_{l=1}^{L-2}(I-P_{l,l+2})\nonumber \\
(1-4g)q^Z_1 + gq^Z_2 + (1-4g)q^Z_L + gq^Z_{L-1}.\label{2dil}\eeqs As
was the case in the analysis of single trace operators, we do not
expect the two loop spin chain to be exactly integrable. However, in
the analysis of closed spin chains, it was possible to establish a
perturbative integrablity of the large $N$  dilatation operator. In
other words, it was possible to construct a Bethe ansatz for the
eigenstates of the two loop Hamiltonian which was accurate to
$O(g)$\cite{paba}. It is thus reasonable to examine if the two loop
large $N$ dilatation operator exhibits this perturbative notion of
integrability in the background of giant gravitons.

The method of the perturbative assymptotic Bethe ansatz (PABA)
advocated by Staudacher in\cite{paba} provides one with a very
direct and transparent method for checking whether or not a given
spin chain, whose Hamiltonian can be expressed as a power series in
some coupling constant (in our case $g$), admits of a Bethe ansatz
order by order in perturbation theory. This method was successfully
applied to the study of the mixing of single trace $su(2)$ operators
of the gauge theory up to three loops in perturbation theory. In the
sector of single trace operators, the large $N$ dilatation operator
maps to closed spin chains with periodic boundary conditions. For
such spin chains, there are no boundaries, and the magnons only
scatter off each other. Establishing the existence of a PABA for the
closed spin chains  thus consists of a two step process.\\{\bf( 1:)}
Finding the  two body $S$ matrix for two magnon scattering while
allowing for the $S$ matrix to have an appropriate $g$ dependence
such that the two magnon Bethe state is an eigenstate of the
Hamiltonian in the perturbative sense mentioned above.\\{\bf(
2:)}Generalizing the two body solution to a solution of the many
body problem using the kind of reasoning used at the end of the
previous section, which basically amounts to assuming factorized
scattering for the problem.\\
For the problem at hand, the only difference between the two loop
Hamiltonian(\ref{2dil}) and the corresponding spin chains describing
the two loop dilatation operator for single trace states lies in the
boundary conditions and the boundary terms(the second line in
(\ref{2dil})). Thus, when the magnons are far away from the
boundaries, the scattering between the magnons is described by the
same $S$ matrix that appears in the case of the closed spin chains,
which was already found in\cite{paba}. Thus, for the purposes of the
present analysis, we can borrow from\cite{paba} the formula for the
two loop bulk two magnon scattering matrix.  However, for the
present problem, we also need to perform a PABA to account for the
scattering of the magnons from the boundaries i.e compute the
boundary scattering matrix accurate to  two loop order. As in the
case of the one-loop analysis performed earlier in the paper, the
boundary scattering matrix, will be found by solving the one magnon
problem. To  clinch the issue of perturbative integrabilty in
analysis  of the open spin chain would require a demonstration that
the Bethe state formed by merging the two asymptotic solutions is an
eigenstate of the Hamiltonian. To put it differently, we shall have
to check that two magnon state that has as the bulk $S$ matrix the
one found in\cite{paba}, and as the boundary $S$ matrix the
appropriate solution of the one magnon problem is an eigenstate of
the Hamiltonian. This unfortunately is not the case. We shall now
proceed to substantiate these comments by explicit computations in
the following sections.

\subsection{One Magnon Problem:}As in the one-loop case, we shall
first compute the boundary $S$ matrix of the spin chain by solving
the one-magnon problem. An important distinction from the one-loop
analysis is that the Bethe wave function is {\it not} simply a
superposition of plane waves. We shall also have to allow for some
exponential decay of the wave function around the boundaries. The
analysis below can be thought of as the two loop PABA for the one
magnon problem. Although this is ostensibly a one-body problem, the
presence of the boundary induces a two body effect through the
interactions of the impurity with the boundary. Just like the two
body problem in Staudacher's PABA\cite{paba} required the
introduction of an exponential decay of the two body wave function
for the Bethe ansatz to be consistent, the one magnon problem in the
present analysis requires that the one body wave function be allowed
to have an exponential decay around the boundaries. The Bethe ansatz
that incorporates this requisite feature can be summarized in the
following form for the one magnon wave function:
 \beq |\Psi> = \sum_{x=1}^L\Psi (x)|x>\eeq where\beq
\Psi (x) = A(k)e^{ikx} - A(-k)e^{-ikx} + a(k)g^x + b(k)g^{L-x}\eeq
As is evident from the above formula, the third and the fourth terms
account for the exponential decay of the wave function; the strength
of the decay being given by the 't Hooft coupling constant. To avoid
having to write factors involving the momenta, let us call \beqs
A(k)e^{ikx} - A(-k)e^{-ikx}&=&\psi(x)\nonumber \\ a(k) =
\phi(1)\nonumber\\
b(k) =\phi(L)\eeqs As in the one loop analysis, when the impurity is
sufficiently far away from the boundary i.e  for any $2<x<L-2$ we
have the condition:\beq (1-4g)(2\psi(x) - \psi(x-1) - \psi(x+1)) + g
(2\psi(x) - \psi(x-2) - \psi(x+2)) = E\Psi(x)\eeq with\beq E(k) =
4\sin^2\left(\frac{k}{2}\right) - 16g
\sin^4\left(\frac{k}{2}\right)= E_1(k) +
gE_2(k).\label{2lpenergy}\eeq Special care is needed to analyze the
situations $x=2,1,L, L-1$. The coefficient of $|x=2>$, after the
action of the Hamiltonian on the Bethe wave function is\beq
(2-4g)(2\psi(2) - \psi(1) - \psi(3)) + g (2\psi(2) - \psi(4)) -
g\phi(1).\eeq Thus we have the condition\beq \psi(0) = \phi(1)\eeq
Similarly, the coefficient of $|x=1>$ is\beq (1-4g) (2\psi(1) -
\psi(2)) + g(\psi(1) - \psi(3)) + 2g\phi(1)\eeq This leads to the
condition:\beq (1-4g)\psi(0) + g\psi(-1) -g\psi(1) + 2g\phi(1) =
gE_1(k)\phi(1)\eeq where $E_1(k) =4\sin^2\left(\frac{k}{2}\right)$.
There are similar equations that follow from the scattering at the
right boundary. These equations may be written in a way that is at
par with the one loop equations(\ref{comp1}) i.e\beqs A(k)\alpha
(-k) - A(-k)\alpha(k) =0\nonumber\\
A(k)\beta (k) - A(-k)\beta(-k) =0.\eeqs $\alpha $($\beta $)
determine the two loop left (right) boundary scattering amplitudes
and are given by \beqs \alpha(-k) &=&(1-2g) + ge^{-ik} - ge^{ik}
-gE_1(k)\nonumber \\
\beta(k) &=&e^{ik(L+1)}\left((1-2g) + ge^{ik} - ge^{-ik}
-gE_1(k)\right).\label{2loopb}\eeqs As  at the one loop level, the
compatibility conditions $\alpha(k)\beta(k) = \alpha(-k)\beta(-k)$
determine the momenta to be \beq k = \frac{n\pi -\theta
(k)}{L+1},\eeq where\beq \theta (k) = \tan
^{-1}\left[\frac{2g\sin(k)}{(1-2g) - gE_1(k)}\right].\eeq
  It is understood that the expression
for $\theta $ is to be regarded as an approximate expression
accurate to $O(g)$. $\theta(k) $ gives us the two loop correction
from the boundary scattering to the allowed values of momenta. This
completes the Bethe ansatz for the one magnon probelm.
\subsection{Two Magnon Problem:}Having solved the one magnon problem, we can now analyze the case of two magnons.
Once again, we shall have to propose a Bethe ansatz, that accounts
for the exponential decay of the wave functions when the magnons
interact with each other or with the boundary. The appropriate PABA
can be summarized in the following two body wave function \beq
|\Psi> = \sum_{x<y}\Psi(x,y)|x,y> = \sum_{x<y}\psi(x,y)|x,y> +
g\sum_{y>1} \phi(1,y)|1,y> + g\sum_{x<L}
\tilde\phi(x,L)|x,L>.\label{2magnon}\eeq As in Saudacher's
PABA\cite{paba} we shall let $\psi(k_1,k_2)$ be a superposition of
plane waves, with some exponential decay that takes place when the
magnons interact with each other, while $\phi, \tilde \phi $ will
measure the amplitude of decays around the boundaries. More
specifically, we shall have\beqs \psi(x_1,x_2) =
\sum_{p}\sigma(p)e^{i(k_1x_1 + k_2x_2)}\left( A(k_1,k_2) +
B(k_1,k_2)g^{|x_1-x_2|}\right).\eeqs Once again, for two magnons at
$x_1>2$ and $L-2>x_2>x_1+2$, the coefficient of $|x_1,x_2>$ after
the action of the Hamiltonian is\beqs(1-4g)(4\psi(x_1,x_2) -
\psi(x_1-1,x_2)-\psi(x_1+1,x_2) - \psi(x_1,x_2-1)-
\psi(x_1,x_2+1))\nonumber \\
+g(4\psi(x_1,x_2) - \psi(x_1-2,x_2)- \psi(x_1+2,x_2) -
\psi(x_1,x_2-2) - \psi(x_1,x_2+2))\nonumber\\
= E(k_1,k_2)\psi(x_1,x_2)
=\sum_{i=1,2}\left(4\sin^2\left(\frac{k_i}{2}\right) -
16g\sin^4\left(\frac{k_i}{2}\right)\right) \psi(x_1,x_2)\eeqs At the
special situation when $x_2 = x_1+2$, one has the condition\beq
2\psi(x,x+2) - \psi(x+2,x+2) - \psi(x,x) = 0,\label{scat1}\eeq while
when $x_2=x_1+1$, one has the condition\beqs g(\psi(x_1+2,x_2) +
\psi(x_1,x_2-2) - \psi(x_1-1,x_2-1) - \psi(x_1+1,x_2+1))\nonumber \\
(4g-1)(2\psi(x_1,x_2) - \psi(x_1+1,x_2) - \psi(x_1,x_2-1))
=0\label{scat2}\eeqs This part of the computation is the same as the
PABA for periodic spin chains carried out by Staudacher
in\cite{paba}. The above conditions serve to fix the function
$B(k_1,k_2)$ and to determine the bulk scattering matrix. However
the boundary processes have to be treated separately which we shall
now proceed to do. Now, for any $x_2$, the condition obtained from
looking at the coefficient of $|2,x_2>$ is\beq \psi(0,x_2) =
\phi(1,x_2).\label{scat3}\eeq One also has a similar condition at
the right boundary; namely\beq \psi(x_1,L+1) =
\phi(x_1,L).\label{scat4}\eeq Furthermore, for any $x_2>2$ and
$x_1<L-1$ the coefficients of $|1,x_2>$ and $|x_1,L>$ lead to
following two boundary scattering equations respectively.\beqs
A(k_1,k_2)\alpha (-k_1) -
A(-k_1,k_2)\alpha (k_1) =0\nonumber \\
A(k_1,k_2)\beta (k_2) - A(k_1,-k_2)\beta (-k_2),\label{scat5}\eeqs
with $\alpha,
\beta $ given by(\ref{2loopb}).
Equations(\ref{scat1}, \ref{scat2}, \ref{scat3}, \ref{scat4},
\ref{scat5}) completely fix all the undertermined functions in the
problem. Indeed, they also lead to the Bethe equations for the
momenta \beq\frac{\alpha (k_i)\beta (k_i)}{\alpha (-k_i)\beta(-k_i)}
= \prod_{j\neq i}S(-k_i,k_j)S(k_j,k_i).\label{2lpba}\eeq $\alpha $
and $\beta $  are given by (\ref{2loopb}), while the bulk $S$ matrix
is the solution of
(\ref{scat1},\ref{scat2},\ref{scat3},\ref{scat4}). It is the same as
the one  presented in\cite{paba,bs-all-loops,hubbard}.\beq S(p,p') =
\frac{\phi(p) - \phi(p') +i}{\phi(p) - \phi(p') -i},\eeq with the
phase shift (thought of as an expression accurate to order $\lambda$
for the present problem) being given by \beq \phi(p) =
\frac{1}{2}\cot\left(\frac{p}{2}\right)\sqrt{1+8\lambda
\sin^2\left(\frac{p^2}{2}\right)}.\eeq

This situation is very much at par with what one had at one
loop(\ref{one-loop}). The boundary interactions so far have not
introduced any three-body interactions. The effect of the left
(right) boundary scattering on $\psi(x_1,x_2)$ produces the same
algebraic equations  satisfied by $\psi(x)$ in the one magnon
problem. Thus in the analysis so far, the boundary has not 'seen'
the second particle. That is not surprising as we have not analyzed
the only situation that can lead to a three body process, which is
when both the impurities are near the boundary: i.e we need to study
the coefficient of the state $|1,2>$ after the action of the
Hamiltonian on the Bethe wavefunction. This is the only scenario (at
this loop order) in which all the three 'particles', the two
impurities and the boundary can interact with each other. Thus the
spin chain will be perturbatively integrable, if the coefficient of
$|1,2>$ does not lead to any new conditions. Unfortunately, that is
not the case. It is a straightforward exercise to see that the
coefficient of $|1,2>$ leads to the condition  \beq
\sum_{p}\sigma(p)A(k_1,k_2)\tilde \alpha
(-k_1,k_2)=0,\label{leftviol}\eeq where \beq
\tilde{\alpha}(-k_1,k_2) =e^{ik_2}\left((1-4g) +g\left( -e^{ik_1}
+2e^{-ik_2} + e^{-ik_1}-E_1(k_1)\right)\right)\eeq Similarly, at the
right boundary, the coefficient of $|L-1,L>$ leads to\beq
\sum_p\sigma(p)A(k_1,k_2)\tilde \beta
(k_1,k_2)=0,\label{rightviol}\eeq where\beq \tilde \beta(k_1,k_2) =
e^{i(k_1(L-1)+k_2(L+1))}\left((1-4g) + g (e^{ik_2} - e^{-ik_2}
+2e^{ik_1}-E_1(k_2))\right)\eeq

These are  new conditions which imply that  the Bethe state is {\it
not} an eigenstate of the Hamiltonian even after one finds bulk and
boundary S matrices which are solutions of the two body Schrodinger
equations.  The presence of both the momenta $k_1,k_2$ in boundary
scattering phase shifts $\tilde{\alpha}, \tilde \beta $ in the above
equations is the signature of three body interactions that prevent
the separation of the scattering matrix into a bulk and a boundary
part. The extra three body interaction coming from the special
possibility of both the particles being close to the boundary thus
shows us that the conventional Bethe ansatz does not quite work in
the case of the present problem.

Though the problem does not appear to be integrable, at least from
the PABA point of view, we can nevertheless use the Bethe equations
to construct some exact two particle eigenstates.  This can be
carried out  by allowing for a linear superposition of the the Bethe
wave functions of the kind discussed above with various values of
the momenta. Instead of(\ref{2magnon}), we can look for a wave
function of the type\beq\Psi(x,y) = \Psi(x,y) +\Gamma \Psi ^\prime
(x,y) + \Delta \Psi^{\prime \prime}(x,y),\eeq where $\Psi ^\prime,
\Psi^{\prime \prime}$ have exactly the same functional form as $\Psi
$, but they are chosen to have  different momenta, $k_1^\prime,
k_2^\prime$, and $k_1^{\prime \prime}, k_2^{\prime \prime}$
respectively. The functions are also chosen to satisfy the
scattering
equations(\ref{scat1},\ref{scat2},\ref{scat3},\ref{scat4},\ref{scat5}).
Furthermore, we shall choose the three pairs of momenta to be
related by\beq \sum_{i=1,2}E(k_i) = \sum_{i=1,2}E(k^\prime _i) =
\sum_{i=1,2}E(k^{\prime \prime} _i)\eeq with $E(k)$ given by
(\ref{2lpenergy}). We can now fix the two undertermined functions of
the momenta $\Delta, \Gamma$ by requiring the state to be an
eigenstate of the Hamiltonian. Clearly, the equations that need to
be satisfied by $\Gamma$ and $\Delta $ follow from
(\ref{leftviol},\ref{rightviol}). They are
 \beqs\sum ^\prime A(k_1,k_2)\tilde \alpha (-k_1,k_2) +\Gamma\sum ^\prime\left( A(k^\prime_1,k^\prime_2)\tilde \alpha
(-k^\prime_1,k^\prime_2)\right)+\Delta\sum ^\prime \left(A(k^{\prime
\prime}_1,k^{\prime \prime}_2)\tilde \alpha (-k^{\prime
\prime}_1,k^{\prime \prime}_2)\right)=0\eeqs and \beqs \sum ^\prime
A(k_1,k_2)\tilde \beta (k_1,k_2)+ \Gamma\sum
^\prime\left(A(k^\prime_1,k^\prime_2)\tilde \beta
(k^\prime_1,k^\prime_2)\right)+ \Delta\sum ^\prime\left(A(k^{\prime
\prime}_1,k^{\prime \prime}_2)\tilde \beta (k^{\prime
\prime}_1,k^{\prime \prime}_2)\right)=0\eeqs In the above equations
the 'primed' sums refer to summations over all permutations and
negations of momenta as we have carried out in the rest of the
paper; i.e $\sum ^\prime \equiv \sum_p \sigma(p)$.These equations
can indeed be solved, allowing us to construct some special
eigenstates of the Hamiltonian. In other words, the two body problem
is solvable in a sub-space of the Hilbert space spanned by the Bethe
wavefunctions. In the presence of integrability, every Bethe state
corresponding to any of the allowed values of momenta determined
by(\ref{2lpba}) would have been an eigenstate of the Hamiltonian.
Though that is not the case, special linear combinations of the
Bethe states do still solve the two magnon problem. It is however
not clear to us if the above two body solutions can be utilized to
solve the $n>2$ magnon problem.

\section{Epilogue:}
The test of integrability carried out in the above analysis differs
significantly from the usual one carried out to determine whether or
not particular boundary conditions maintain the integrability of a
given integrable spin chain. One usually declares a spin chain with
open boundary conditions to be integrable if its bulk and boundary
$S$ matrices satisfy the boundary Yang-Baxter or reflection
algebra\cite{open-3}. This, for example, is the notion of
integrability used in\cite{ber-1} for the open $so(6)$ spin chain.
However, it has to be kept in mind that the $S$ matrices satisfying
the reflection algebra is a necessary and \textit{not} a sufficient
condition for integrability. Moreover, the bulk and boundary $S$
matrices truly become matrices when the impurities carry some
internal degrees of freedom, which is not the case in the present
problem, where the impurity can only be of one kind. The reflection
algebras too, become non-trivial matrix relations only when the
scattering matrices describe particles with internal degrees of
freedom, and hence become non-commutative matrices. In a two
component system like the present one, where the scattering matrices
are just numbers depending on the momenta, the  reflection equations
are trivially satisfied. Thus, the analysis presented above also
gives us a curious example of a situation where the necessary
condition for integrability, namely the associated reflection
algebra, is satisfied, while there does not appear to be a Bethe
ansztz for the problem.

It is also instructive to compare the present study with those
concerning the Inozemtsev spin chain. As mentioned before, the two
(and three) loop $su(2)$ dilatation operator in the sector of single
trace operators has been shown to be embedded in a particular
integrable long ranged spin chain known as the Inozemtsev spin
chain. The Inozemtsev chain has a free parameter in which the
Hamiltonian can be expanded and it is part of a complete set of
commuting charges. It has been shown in\cite{serban-3lp}that it is
possible to fine tune this free parameter so that up to the three
loop order, the $su(2)$ dilatation operator and the Inozemtsev spin
chain become identical. Indeed, this is also the easiest way to
establish perturbative integrability up to the three loop order in
the $su(2)$ sector of the gauge theory. However, the picture changes
when one considers non-periodic boundary conditions. To the best of
our knowledge, the only boundary conditions other than the periodic
one that are compatible with the integrability of the Inozemntsev
and other Inozemtsev like  long ranged spin chains such as the
celebrated Haldane Shastry model are special reflecting boundary
conditions, where the spin chain interacts with its image on the
other sides of the left and right boundaries in very specific ways,
see for example\cite{xing-open, bn-open, serban-open}. The two loop
(bulk) dilatation operator that we have in the present problem is
the same one that arises in the analysis of the closed spin chains,
and thus, it can be embedded in the Inozemtsev spin chain to this
loop order. The boundary conditions  in the present problem though
are not of the reflecting type that lead to integrable 'open'
versions of the Inozemtsev  or Haldane Shastry spin chains which can
possibly be an explanation for the absence of a consistent Bethe
ansatz for the problem.

It is probably worth reiterating at this juncture that the two loop
analysis presented in this paper is not the only example of  the
emergence of boundary conditions that lead to a potential violation
of  integrability of the dilatation operator of $\cal{N}$=4 SYM. As
mentioned earlier in the paper, the dynamical boundary conditions
generated from the study of spin chains coupled to sub-determinant
like operators in\cite{ber-2} produce quantum spin chains whose
status, at least as far as integrability is concerned, is not clear.
Another related scenario of a violation of integrability in this
gauge theory arises when one studies the large $N$ limit of the
$so(6)$ invariant matrix model obtained as the dimensional reduction
of the scalar sector of the super Yang-Mills theory to one (time)
dimension. The quantization of the large $N$ limit of this matrix
model was carried out recently by Klose in\cite{thomas-1}, and a
breakdown of integrability of the model at higher loops was
explicitly demonstrated.

As a final observation, it might be worth comparing the present
analysis to the one performed in the context of defect conformal
field theories. $\cal{N}$ =2 defect conformal field theories are the
conjectured gauge theory duals for the open string theory generated
by wrapping a $D5$ brane on an $AdS_4\times S^2$ subspace of
$AdS_5\times S^5$. For the $su(2)$ sector of such gauge theories,
the open string degrees of freedom are generated by placing quark
like fields at the ends of the spin chain, which result in open/free
boundary
conditions\cite{defect-1,defect-ch,defect-2,defect-3,defect-4,defect-5,defect-6,defect-7,defect-8,defect-9,defect-10}.
These are different from the boundary conditions that we encountered
in the present work, but they are similar. For such open spin
chains, an all loop Bethe ansatz was recently proposed
in\cite{tristan-1}. Although, these are two different theories, and
it probably does not make sense to compare their spectra, one can
nevertheless compare the generic structure of the Bethe equation
obtained in that analysis to the obstacles encountered in the
present work. In the Bethe ansatz proposed in\cite{tristan-1}, the
boundary scattering matrix did not have any dependence on the 't
Hooft coupling of the gauge theory. Although the present analysis
prevents us from writing down a set of Bethe equations of the
present problem, we were able to determine the boundary scattering
matrix at the two loop level, and found it to have a non-trivial
dependence on the 't Hooft coupling. From the present analysis, it
appears that such dependence is a generic feature of boundary
scattering for (long-ranged) open spin chains, in the absence of
reflective boundary conditions(see for example\cite{bn-open,
serban-open}). Thus it might be a reasonable goal to perform an
explicit two loop computation for the $su(2)$ sector of $\cal{N}$ =2
defect conformal field theories, as there might be hidden surprises
to be uncovered. At the very least, it might be possible to
understand their integrability by studying the relation between the
boundary conditions that emerge out of the defect conformal field
theories and the reflecting boundary conditions that have been
studied in the context of long-ranged spin chains like the
Haldane-Shastry and Inozemtsev chains in the past\cite{xing-open,
bn-open, serban-open}.

Thus in summary, the analysis presented in this paper presents some
 evidence in favor of  lack of higher loop integrability in
$\cal{N}$ =4 SYM when one ventures into the study of the gauge
theory dual of open string like operators. It is worth noting that
similar obstacles in the direction of the implementation of Bethe
ansatz methods at higher loops in the open string sector of this
gauge theory were also arrived at recently in \cite{yosh-h}. The
present analysis of integrability was carried out in a somewhat
unconventional approach. It would perhaps be worthwhile to carry out
a numerical study of the spectrum of the two loop spin chain
presented in the paper along the lines of the involved work done
in\cite{virial}. The empirical knowledge of the spectrum can then be
used to study whether or not the degeneracies responsible for the
one loop integrability are lifted at two loops and possibly provide
a more direct study of integrable aspects of the problem. A
numerical investigation of the spectrum of the gauge thoery dual of
open strings coupled to non-maximal giant gravitons was recently
carried out in\cite{ber-recent}. In that paper some tell-tale
numerical signs of integrability were also presented. Thus, it would
be extremely interesting to understand if integrability is present
in the problem in some way that is independent of the existence of a
Bethe ansatz.

\vskip 0.3cm \noindent {\bf Acknowledgements:} It is a great
pleasure to thank Niklas Beisert, Dimitra Karabali, Tristan
McLoughlin, Parameswaran Nair, Alexios Polychronakos, Sarada Rajeev,
Radu Roiban and Matthias Staudacher for valuable discussions at
various stages of this work. Special thanks are also due  to Niklas
Beisert, Tristan McLoughlin, Parameswaran Nair and Alexios
Polychronakos for their comments on an earlier version of this
manuscript.

\end{document}